\documentclass[prl,twocolumn, superscriptaddress,showpacs,preprintnumbers,amsmath,amssymb]{revtex4}
\usepackage{amssymb}
\usepackage{graphicx}
\usepackage{dcolumn}
\usepackage{bm}
\usepackage{amsmath}

\begin{document}

\preprint{NEMS 22 Jan 2010}
\title{Damping in high-frequency metallic nanomechanical resonators}
\author{F. Hoehne}
\email[Electronic address: ]{hoehne@wsi.tum.de} \affiliation{Walter Schottky Institut, Technische
Universit\"{a}t M\"{u}nchen, Am Coulombwall 3, 85748 Garching, Germany} \affiliation{NEC Nano Electronics
Research Laboratories and RIKEN Advanced Science Institute, Tsukuba, Ibaraki 305-8501, Japan}

\author{Yu.\ A. Pashkin}
\altaffiliation[On leave from ]{Lebedev Physical Institute, Moscow 119991, Russia} \affiliation{NEC Nano
Electronics Research Laboratories and RIKEN Advanced Science Institute, Tsukuba, Ibaraki 305-8501, Japan}

\author{O. Astafiev}
\affiliation{NEC Nano Electronics Research Laboratories and RIKEN Advanced Science Institute, Tsukuba,
Ibaraki 305-8501, Japan}

\author{L. Faoro}
\affiliation{Physics Department, Rutgers, 136 Frelinghuysen Rd, Piscataway, NJ 08854 USA} \affiliation{LPTHE,
CNRS UMR 7589, 4 place Jussieu, Paris, France}

\author{L. B. Ioffe}
\affiliation{Physics Department, Rutgers, 136 Frelinghuysen Rd, Piscataway, NJ 08854 USA}

\author{Y. Nakamura}
\affiliation{NEC Nano Electronics Research Laboratories and RIKEN Advanced Science Institute, Tsukuba,
Ibaraki 305-8501, Japan}

\author{J. S. Tsai}
\affiliation{NEC Nano Electronics Research Laboratories and RIKEN Advanced Science Institute, Tsukuba,
Ibaraki 305-8501, Japan}

\date{\today}

\begin{abstract}
We have studied damping in polycrystalline Al nanomechanical resonators by measuring the temperature
dependence of their resonance frequency and quality factor over a temperature range of 0.1 -- 4~K. Two
regimes are clearly distinguished with a crossover temperature of 1~K. Below 1~K we observe a logarithmic
temperature dependence of the frequency and linear dependence of damping that cannot be explained by the
existing standard models. We attribute these phenomena to the effect of the two-level systems characterized
by the unexpectedly long (at least two orders of magnitude longer) relaxation times and discuss possible
microscopic models for such systems. We conclude that the dynamics of the two-level systems is dominated by
their interaction with one-dimensional phonon modes of the resonators.
\end{abstract}

\maketitle

Nanoelectromechanical systems have recently attained a lot of interest due to a variety of promising
applications such as ultra-sensitive mass measurement~\cite{Mamin01} and single-spin detection~\cite{Rugar04}
as well as due to their suitability for doing quantum-limited measurements~\cite{LaHaye,Teufel,Rocheleau}.
For this, nanomechanical resonators with high resonance frequencies in combination with high quality factors
are required. So far, nanomechanical resonators mostly made of single crystal semiconductor and dielectric
materials have been studied, and frequencies above 1~GHz with quality factors of about 500 have been
achieved~\cite{Huang03}. Low-temperature studies on such resonators reveal glass-like
behavior~\cite{Zolfagharkhani05,Imboden09} and suggest that this may be due to the interaction of the
flexural beam modes with two-level-systems (TLS). Metallic beams can be fabricated using technologies known
for metallic nanoelectronic devices, which therefore allows easy integration of mechanical degrees of freedom
into such devices. Although fabrication methods for metallic beams of nanoscale size have been developed
recently~\cite{Husain03,Li08}, very little is known about loss mechanisms in such beams. In this respect,
understanding damping in nanoscale mechanical resonators is of primary importance. In this work we
demonstrate that the quality factor of Al doubly clamped beams at mK temperatures is affected by two-level
systems. The observed temperature dependences of damping and resonance frequency are characteristic for
amorphous materials with TLS relaxation due to one dimensional phonons. Such a dependence cannot be
interpreted within the standard TLS model~ \cite{Hunklinger1986,Phillips87}, and we propose an alternative
explanation. The apparent similarity between amorphous insulators and polycrystalline metals and their
dramatic difference from amorphous metals implies a different nature of TLS in metallic beams.
\begin{figure}[bp]
\includegraphics[width=8cm]{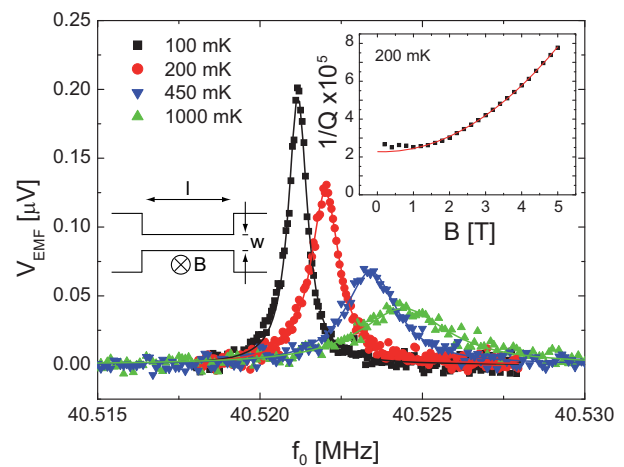}
\caption{(color online) Amplitude of the induced electromotive force of the 5~$\protect\mu $m beam at
$B$~=~0.5~T for different temperatures (symbols). The solid lines are Lorentzian fits. Right inset shows
damping 1/$Q$ of the same beam as a function of the magnetic field (black squares). The $B^{2}$-dependence
expected from magnetomotive damping is shown as a fit (red line) to the experimental data. Left inset
presents a layout of the beams studied.} \label{fig:PaperTraces}
\end{figure}
\newline
\indent Doubly clamped beams are fabricated on an oxidized silicon substrate using a trilayer resist
structure with an extra sacrificial calixarene layer~\cite{Li07,Li08}. The beam mask in Ge layer is defined
by electron-beam lithography and reactive ion etching. After metallization and lift-off process, the beams
are suspended by removing the underlying calixarene layer in an oxygen plasma. The Al polycrystalline beams,
with length $l$, width $w$ and thickness $t$ (see left inset of Fig.~\ref{fig:PaperTraces}), are connected to
the central line of prefabricated gold-patterned on-chip coplanar waveguides which are ribbon-bonded to
coaxial lines. The samples are mounted in vacuum space inside the bore of a superconducting solenoid
providing a transverse magnetic field of up to 5~T. All measurements were done in a dilution refrigerator
with a base temperature down to about 50~mK.

We characterize our beams using a conventional magnetomotive measurement scheme~\cite{Cleland96}. The RF
signal from the output of the network analyzer is fed into the coaxial line at the top of the cryostat and
delivered to the chip through a 20~dB attenuator at the 4 K stage. AC current flowing through the beam in a
perpendicular external magnetic field $B$ actuates the device due to the Lorentz force. On resonance the beam
dissipates energy producing a dip in the transmitted signal which is delivered via a second coaxial line to
the room-temperature preamplifier and then to the input of the network analyzer. The power applied to the
beams is low enough to keep them in the linear regime, as seen from the resonances depicted in
Fig.~\ref{fig:PaperTraces} for the 5-$\mu$m long beam. Electromotive force
induced on the beam is estimated from the measured transmission as $V_{\rm EMF}=%
\sqrt{PZ}$, where $P$ is the power difference in the transmitted signal on and off resonance and $Z=50~\Omega
$ is the impedance of the high-frequency line. The resonance frequency $f_{0}$ and damping $1/Q$ are
extracted from these traces by fitting them with a Lorentzian. Damping $1/Q~=~{\Delta } f/f_{0}$ is defined
as a full width at half maximum divided by the center frequency of the Lorentzian fit to the transmitted
power. Damping dependence on the applied magnetic field is shown in the right inset of Fig.~\ref
{fig:PaperTraces}. Magnetomotive damping is proportional to the motional impedance that scales as $B^2$,
therefore it dominates in high magnetic fields and produces an overall parabolic dependence. This is
confirmed by the parabolic fit to the experimental data between 1~T and 5~T shown by the red line and
extrapolated to lower fields. Below about 1~T, damping saturates and deviates from the parabolic dependence,
which may be attributed to the contribution from the field dependent losses in the external circuit.

\begin{figure}[tbp]
\includegraphics[width=8cm]{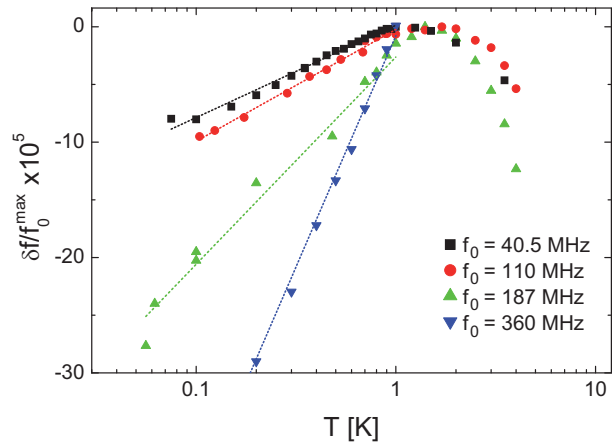}
\caption{(color online) Relative shift of the resonance frequency as a
function of the temperature for different resonators (symbols). For each
device the frequency shift is normalized to the maximum resonance frequency $%
f_{0}^{\mathrm{max}}$. The dotted lines are logarithmic fits to the measured data used to determine the
values for $C$ in Table~\ref{fig:Table_Fit}.} \label{fig:PaperDeltafvsT}
\end{figure}

Figure~\ref{fig:PaperDeltafvsT} shows the relative change of the resonance frequency $\delta
f/f_{0}^{\mathrm{max}}$, for four beams of different dimensions, as a function of temperature on a
logarithmic scale. For each trace, the frequency change is normalized to its maximum resonance frequency
$f_{0}^{\mathrm{max}}$. All the beams exhibit qualitatively the same overall temperature dependence: the
resonance frequency reaches a maximum at about 1.5~K and decreases at higher and lower temperatures.
Moreover, below this temperature, the frequency decreases logarithmically down to the lowest temperature that
could be reached in our measurements, for all four beams, although their slopes are different. It should be
noted here that the beam resonance frequencies change with temperature due to tension created due to the
difference in coefficients of thermal expansion of Al and Si, however, this dependence is monotonous in the
temperature range from 300~K down to about 20~K. Below 20~K the thermal expansion coefficients saturate and
hence no frequency change is expected. Therefore, the characteristic frequency dependence at low temperature
must be attributed to a different mechanism as discussed below.

\begin{figure}[tbp]
\includegraphics[width=8cm]{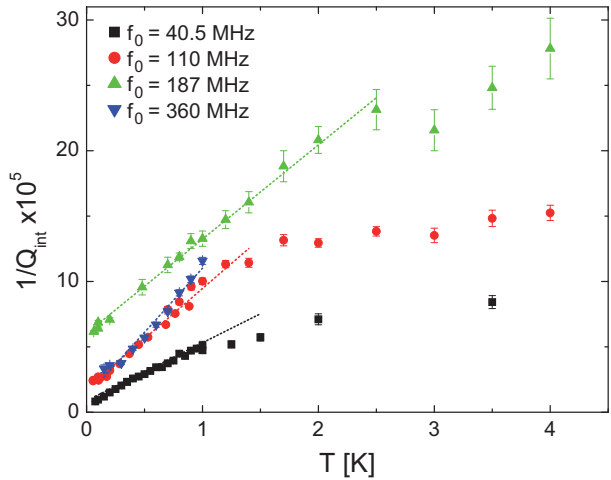}
\caption{(color online) Damping $1/Q_{\mathrm{int}}$ as a function of temperature for four resonators of
different lengths (symbols). Damping increases linearly up to $0.7~-~1.5$~K followed by a weaker temperature
dependence above this temperature. The dotted lines are linear fits to the measured data used to determine
the values for $\nu_F U$.} \label{fig:Paper1QvsT}
\end{figure}

Figure~\ref{fig:Paper1QvsT} shows damping as a function of temperature for the same range as in
Fig.~\ref{fig:PaperDeltafvsT}, but with a linear temperature scale. Again, all four beams show qualitatively
the same temperature dependence. Damping increases linearly with temperature up to about 1~K. Above this
temperature it continues to increase but with a significantly smaller slope.

Now we turn to the analysis of possible loss mechanisms in Al beams. Damping $1/Q$ can be divided into
internal and external as $1/Q~=~1/Q_{\mathrm{int}}~+~1/Q_{\mathrm{ext}}$. For the 5-$\mu $m long beam, the
observed parabolic dependence of damping on magnetic field (see inset of Fig.~\ref{fig:PaperTraces}) is
related to external losses $1/Q_{\mathrm{ext}}$ in the measurement circuit, called magnetomotive
damping~\cite{Cleland99}. Since we are interested in the intrinsic or material-dependent loss mechanisms $
1/Q_{\mathrm{int}}$, we subtract the effect of magnetomotive damping from the measured values of $1/Q$ for
all beams. In addition, since magnetomotive damping scales as $\propto~l^3$, all measurements of the
temperature dependence for the 5-$\mu $m long beam have been performed at 0.5~T.

Several intrinsic mechanisms discussed in the literature can contribute to the dissipation in nanomechanical
resonators at low temperatures. Thermoelastic damping has been shown to be negligible at high frequencies and
low temperatures and therefore does not play any role in the measurements reported here~\cite{Lifshitz00}. In
metals, electron-phonon scattering may also contribute to damping~\cite{Pippard65}, but its contribution in
polycrystalline metals can usually be neglected due to the short mean free path of
electrons~\cite{Loehneysen81}.
%
\newline
\indent Clamping loss is another contribution to damping of the beams. It depends on the beam aspect ratio as
$(w/l)^3$ for in-plane displacement~\cite {Cross01}, therefore this loss mechanism becomes significant for
the beams with a high resonance frequency, i.e., small aspect ratio, as can be seen in
Table~\ref{fig:Table_Fit}. However, this damping mechanism is not expected to show any temperature dependence
and therefore contributes to the data shown in Fig.~\ref{fig:Paper1QvsT} only as a constant background
without affecting the slope. Ideally, the temperature dependence of $1/Q_{\mathrm{int}}$ should be shifted up
for shorter beams provided only their length is changed. Such a tendency is observed in
Fig.~\ref{fig:Paper1QvsT} for the three beams with the same width and thickness (see data for beams with $f_0
$= 40.5, 110 and 187~MHz). Table I shows that with the fabrication method utilized, a resonance frequency
as high as 770~MHz together with a quality factor of 2400 have already been achieved. By optimizing the
beam dimensions, it should also be possible to push the resonance frequency up to above 1~GHz while
preserving a reasonably high quality factor.
%
\begin{table}[tbp]
\caption{Physical parameters of Al resonators with different dimensions. The beam thickness is 0.2~$\mu$m
except for the narrow beams ($w=0.06$~$\mu$m) whose thickness is 0.1~$\mu$m. The resonance frequency $f_0$ and
the quality factor $Q$ are extracted at $ T~=~100$~mK.}
\label{fig:Table_Fit}%
\begin{tabular}{cccccc}
\hline \hline $l \times w$ [$\mu$m$^2$] & 5$\times$0.1 & 3$\times$0.1 & 2$\times$0.1 & 1$\times$0.06 &
0.6$\times$0.06 \\ \hline
$f_0$ [MHz] & 40.5 & 110 & 187 & 360 & 770 \\
$Q$ & 120000 & 41000 & 16000 & 30000 & 2400 \\ $C$ & 3$\times$10$^{-5}$ & 4$\times$10$^{-5}$
&8$\times$10$^{-5}$ & 17$\times$10$^{-5}$ & X \\
$\nu_F U$ & 0.05 & 0.09 & 0.08 & 0.09 & X \\
\hline \hline
\end{tabular}%
\end{table}

%
\indent Qualitatively, the temperature dependence of the resonance frequency and damping in our experiments
looks similar to those observed in amorphous insulators where they are attributed to two-level
systems~\cite{Hunklinger1986}. In these materials the temperature dependence of both the sound velocity and
damping display two regimes with a crossover temperature around $T^{\ast }\sim 1$~K. The low-temperature
regime is characterized by a logarithmic increase of the sound velocity with temperature while the quality
factor changes as a power law with temperature. The crossover between the two regimes occurs when the TLS
relaxation rate $\Gamma$ becomes approximately equal to the measurement frequency $\omega=2\pi f$, which is
close to the beam resonance frequency in our experiments. One expects that in amorphous metals the
interaction of the TLS with electrons is much stronger leading to a much larger value of $\Gamma$, so that
the high-frequency (low-temperature) regime is very difficult to observe, in agreement with the existing
data~\cite{Black79, Esquinazi2002}. The situation in polycrystalline metals, such as the samples studied in
this work, is less clear. In particular, the naive expectation that the defects in these materials should
behave similarly to the ones in amorphous metals was not confirmed experimentally. Instead, surprisingly,
they behave more like defects in amorphous insulators~\cite{Esquinazi1992}. Our data also confirm this fact
and imply that the interaction of the defects with conduction electrons is very small, in agreement with
Ref.~\cite{Konig1995}.

For all mechanisms one expects that the sound velocity changes logarithmically in the low
temperature regime:
\begin{equation}
\frac{\delta f}{f_0}=C\ln \left( \frac{T}{T_{0}}\right)  \label{delta_f},
\end{equation}%
where $C$ is a dimensionless parameter that characterizes the TLS interaction with sound waves and with each
other: $C=\nu _{T}\gamma ^{2}/E$, where $\nu _{T}$ is the TLS density of states, $\gamma $ is the interaction
constant and $E$ is the Young's modulus. Remarkably, the value of $C$ is known to be almost universal for all
amorphous materials, $C~\sim~10^{-3}$~--~$10^{-4}$~\cite{YuLegget1988}. Our measurements on polycrystalline
Al samples give values of $C \approx~0.4$~--~$1.7\times10^{-4}$, consistent with this phenomenology.

In contrast, temperature dependence of damping is very sensitive to the physics of TLS because it directly
probes the TLS relaxation rate:\

\begin{equation}
1/Q(T)\approx \left\{
\begin{array}{cl}
C\displaystyle{\frac{\Gamma }{\omega }} & \quad \omega >\Gamma (T) \\
\\
C & \quad \omega <\Gamma (T)%
\end{array}%
\right.  \label{1/Q}
\end{equation}

The observed linear temperature dependence of damping below 1~K implies that all data below this temperature
correspond to the high-frequency regime $\omega >\Gamma (T)$ and that $\Gamma \propto T$. For this to be
true, even the lowest resonance frequency $f_0\sim 40$~MHz must be larger than the relaxation rate at $T\sim
1$~K. Note that the temperature dependent part of damping $1/Q_{\mathrm{int}}(T)-$
$1/Q_{\mathrm{int}}(0)\lesssim 0.5\times 10^{-4}$ is consistent with the regime $\omega >\Gamma (T)$.

The conclusion $\omega >\Gamma (T)$ is difficult to reconcile with the electron mechanism of TLS relaxation
because for this mechanism $\hbar\Gamma =2\pi (\nu _{F}U)^{2}k_BT$, where $\nu _{F}$ is the electron density
of states and $U$ is their interaction with TLS. Assuming that this condition holds for $ f_0=40$~MHz and
$T=1$~K one would conclude that $\nu _{F}U\lesssim 0.01$, which is much smaller than one expects and observes
for a conventional TLS in a metal, $\nu _{F}U=0.1$ -- $1$ \cite{Hunklinger1986}. The different temperature
dependence of damping studied here from that reported for bulk polycrystalline Al samples also points towards
the phonon dominated relaxation.

This conclusion, however, is in a perfect agreement with the phonon mechanism of TLS relaxation when one
takes into account the fact that for temperatures $T<1$~K the phonon wavelength in Al is $\lambda
>0.25~\mu$m, so that at these temperatures the studied beams are essentially one dimensional structures. The
linear phonon spectrum \cite{LinearSpectrum} implies a constant density of states in a 1D system, and
therefore the TLS relaxation rate is expected (cf. \cite{Hunklinger1986,Phillips87,Guinea2008}) to be
$\hbar\Gamma \sim (a^{2}/wt)k_BT$, where $a$ is the lattice constant. The ratio $(a^{2}/wt)\sim 10^{-5}$
naturally leads to the right order of magnitude of the relaxation rate and its linear temperature dependence.
Unlike the phonon properties, however, the reduced dimensionality of our resonators does not affect their
electron properties or interaction with TLS.

The apparent absence of the TLS-electron interaction in polycrystalline materials points to a different
origin of the TLS in such materials. In amorphous insulators and metals, the TLS are likely to be single
atoms that tunnel atomic distances between two positions. In crystals, a more likely origin are kinks on
dislocations \cite{Hikata}. These kinks are very smooth objects in soft metals (such as Au and Al) due to a
small value of the Peierls barrier and thus may interact very weakly with the electrons \cite{Hirth}.

The conclusions reached above are based on the linear temperature dependence of $1/Q_{\mathrm{int}}(T)$ and
constant $C$ expected for conventional TLS. These assumptions must be reexamined for TLS originating from
smooth kinks on dislocations. For instance, interaction between the kinks in strained samples may not be so
small as between TLS because the external strain creates kinks until the interaction becomes sufficient to
balance the strain. Such interaction may suppress the density of states at low energies similar to spin glass
physics~\cite{MydoshBook}. However, a typical model predicting suppression of the density of states and a
strong linear $T$-dependence damping would also give a sound velocity change different from
Eq.~(\ref{delta_f}).

Observation of a single TLS and study of its dynamics in polycrystalline materials would be the most direct
way to identify mechanisms discussed above. Alternatively, one can probe the interaction with electrons by
studying the effect of superconductivity on the TLS relaxation rate and $1/Q_{\mathrm{int}}(T)$.

In conclusion we have reported measurements of the temperature dependence of damping and the resonance
frequency of the fundamental mode of doubly clamped metallic nanomechanical resonators. Our data indicate
that these are dominated by unconventional TLS with a long relaxation time which can be associated with
dislocation kinks.

We would like to thank T.F. Li and H. Im for technical assistance.
This work was supported by CREST-JST and MEXT kakenhi "Quantum Cybernetics". LF and LBI were supported by ARO
W911NF-06-1-0208 and DARPA HR0011-09-1-0009. \newline


\begin{thebibliography}{99}
\bibitem{Mamin01} H. J. Mamin and D. Rugar, Appl. Phys. Lett. \textbf{79},
3358 (2001).

\bibitem{Rugar04} D. Rugar, R. Budakian, H. J. Mamin and B. W. Chui, Nature
\textbf{430}, 329 (2004).

\bibitem{LaHaye}M. D. LaHaye, O. Buu, B. Camarota and K. C. Schwab, Science \textbf{304}, 74 (2004).

\bibitem{Teufel}J. D. Teufel , T. Donner, M. A. Castellanos-Beltran , J. W. Harlow and K. W. Lehnert,
Nature Nanotechnology \textbf{4}, 820 (2009).

\bibitem{Rocheleau}T. Rocheleau, T. Ndukum, C. Macklin, J. B. Herzberg, A. A. Clerk and
K. C. Schwab, Nature \textbf{463}, 72 (2010).

\bibitem{Huang03} X. M. H. Huang, C. A. Zorman, M. Mehregany and M. L.
Roukes, Nature \textbf{421}, 496 (2003).

\bibitem{Zolfagharkhani05} G. Zolfagharkhani, A. Gaidarzhy, S. B. Shim, R.
L. Badzey, and P. Mohanty, Phys. Rev. B \textbf{72}, 224101 (2005).

\bibitem{Imboden09} M. Imboden and P. Mohanty, Phys. Rev. B \textbf{79},
125424 (2009).

\bibitem{Husain03} A. Husain, J. Hone, H. W. Ch. Postma, X. M. H. Huang,
T. Drake, M. Barbic, A. Scherer, and M. L. Roukes, Appl. Phys. Lett. \textbf{%
83}, 1240 (2003).

\bibitem{Li08} T. F. Li, Yu. A. Pashkin, O. Astafiev, Y. Nakamura, J.~S.
Tsai, and H. Im, Appl. Phys. Lett. \textbf{92}, 043112 (2008).

\bibitem{Hunklinger1986} S. Hunklinger and A.\ K. Raychaudhuri, Prog. Low
Temp. Phys. \textbf{9}, 265 (1986).

\bibitem{Phillips87} W. A. Phillips, Rep. Prog. Phys. \textbf{50}, 1657
(1987).

\bibitem{Li07} T. F. Li, Yu. A. Pashkin, O. Astafiev, Y. Nakamura, J.~S.
Tsai, and H. Im, Appl. Phys. Lett. \textbf{91}, 033107 (2007).

\bibitem{Cleland96} A. N. Cleland and M. L. Roukes, Appl. Phys. Lett.
\textbf{69}, 2653 (1996).

\bibitem{Cleland99} A. N. Cleland and M. L. Roukes, Sens. Actuators A
\textbf{72}, 256 (1999).

\bibitem{Lifshitz00} R. Lifshitz and M. L. Roukes, Phys. Rev. B \textbf{61},
5600 (2000).

\bibitem{Pippard65} A. B. Pippard, \textsl{The Dynamics of Conduction
Electrons} (Gordon and Breach, New York, 1965).

\bibitem{Loehneysen81} H. v. L\"{o}hneysen, Phys. Rep. \textbf{79}, 161 (1981).

\bibitem{Cross01} M. C. Cross and R. Lifshitz, Phys. Rev. B \textbf{64},
085324 (2001).


\bibitem{Black79} J. L. Black and P. Fulde, Phys. Rev. Lett. \textbf{43},
453 (1979).

\bibitem{Esquinazi2002} R. K\"{o}nig, M. A. Ramos, I. Usherov-Marshak, J. Arcas-Guijarro,
A. Hernando-Ma\~{n}eru, and P. Esquinazi,
Phys. Rev. B \textbf{65}, 180201(R) (2002).

\bibitem{Esquinazi1992} P. Esquinazi, R. K\"{o}nig and F. Pobell, Z. Phys. B
\textbf{87}, 305 (1992).

\bibitem{Konig1995} R. K\"{o}nig, P. Esquinazi and B. Neppert, Phys. Rev. B
\textbf{51}, 11424 (1995).

\bibitem{YuLegget1988} C. C. Yu and A. J. Leggett, Comments Cond. Mat. Phys.
\textbf{14}, 231 (1988).

\bibitem{LinearSpectrum} Very long one dimensional beams support the
flexural mode with the $\omega \sim k^{2}$ spectrum. This mode would lead to the $\Gamma \propto \sqrt{T}$
behavior predicted in \cite{Guinea2008}. Our direct measurements of the fundamental resonance frequencies of
the beams listed in Table I show that the spectrum remains roughly linear in the studied samples even for the
lowest frequencies. This is probably due to the high tension in these samples caused by different thermal
contraction of Al and Si. The quadratic spectrum of the flexural mode in the non-stretched sample might be
the origin of the $Q^{-1}(T)\propto \sqrt{T}$ reported for Au beams in Ref. \cite{Venkatesan2009}

\bibitem{Guinea2008} C. Seo\'{a}nez, F. Guinea and A. H. Castro Neto, Phys. Rev.
B \textbf{77}, 125107 (2008).

\bibitem{Venkatesan2009} A.~Venkatesan, K.~J.~Lulla, M.~J.~Patton, A. D. Armour, C. J. Mellor, and J. R. Owers-Bradley,
arxiv:0912.1281 (2009).

\bibitem{Hikata} A. Hikata and C. Elbaum, Phys. Rev. Lett. \textbf{54}, 2418 (1985).

\bibitem{Hirth} J. P. Hirth and J. Lothe, \textsl{Theory of dislocations} (Wiley, New York, 1981).

\bibitem{MydoshBook} J. A. Mydosh, \textsl{Spin glasses: an experimental introduction} (Taylor \&\ Francis, London, 1993).
\end{thebibliography}
\end{document}